\documentclass[aps,prb, twocolumn,showpacs,preprintnumbers,superscriptaddress]{revtex4}

\usepackage{graphicx}
\usepackage{dcolumn}
\usepackage{bm}

\begin{document}

\title{Magnetization and transport properties in the superconducting Pr$_{2}$Ba$_{4}$Cu$_{7}$O$_{15-\delta }$ with metallic double-chain}

\author{Shibuki Toshima}
\author{Michiaki Matsukawa} 
\email{matsukawa@iwate-u.ac.jp }
\author{Taiji Chiba} 
\author{Satoru Kobayashi} 
\affiliation{Department of Materials Science and Engineering, Iwate University , Morioka 020-8551 , Japan }
\author{Sigeki Nimori}
\affiliation{National Institute for Materials Science, Tsukuba 305-0047 ,Japan}

\author{Makoto Hagiwara}
\affiliation{Kyoto Institute of Technology, Kyoto 606-8585,Japan}

\date{\today}

\begin{abstract}
We have reported the effect of pressure on the magnetization, and transport properties in the nominal composition Pr$_{2}$Ba$_{4}$Cu$_{7}$O$_{15-\delta }$ synthesized by a sol-gel technique. 
A reduction treatment of the as-sintered sample in vacuum causes higher superconductivity achieving $T_{c,on}=\sim 30$ K for $\delta =0.94$.  
Application of hydrostatic pressure on the oxygen depleted sample enhances its  onset temperature up to 36 K at 1.2 GPa, indicating the nearly optimum doping level of the charge carrier in comparison to the pressure dependence of lower $T_{c}$ samples with $\delta =0.45$.  Seebeck coefficient of the superconducting sample shows a metallic conduction, followed by  a clear drop below $T_{c,on}$ and is in its temperature dependence below 100 K quite different from that of the non-superconducting one. This finding strongly suggests a dramatic change of the electronic state along the CuO double chain due to the reduction treatment for the appearance of superconductivity .
\end{abstract}

\renewcommand{\figurename}{Fig.}
\maketitle
\section{INTRODUCTION}
Since the discovery of high-$T_{c}$ copper-oxide superconductors,  extensive studies 
on strongly electron correlated system  have been in progress on the basis of physical properties 
of two-dimensional (2D) CuO$_{2}$ planes.  Moreover, from the viewpoint of low-dimensional physics, 
particular attention is paid to the physical role of one-dimensional (1D) CuO chains included 
in some families of high-$T_{c}$ copper oxides such as Y-based superconductors with the transition 
temperature $T_{c}$=$\sim$ 92K .  
It  is well known that the Pr-substitution for  Y-sites in YBa$_{2}$Cu$_{3}$O$_{7-\delta}$ (Y123) 
and  YBa$_{2}$Cu$_{4}$O$_{8}$ (Y124) compounds dramatically suppresses $T_{c}$  and 
superconductivity in CuO$_{2}$ planes disappears beyond the critical value of  Pr , $x_{c}$=0.5 and 0.8, 
respectively. \cite{SO87,HO98} Such a suppression effect due to Pr-substitution on superconductivity 
has been explained in terms of the hybridization model with respect to 
Pr-4$f$ and O-2$p$ orbitals.\cite{FE93}  Y124 compound with double chains is thermally stable up to 
800 $^\circ$C 
,while in Y123/7-$\delta$ oxygen deficiencies are easily introduced at lower annealing temperatures 
.\cite{JO87}  
Intermediate between PrBa$_{2}$Cu$_{3}$O$_{7-\delta}$ 
(Pr123) with single chains and PrBa$_{2}$Cu$_{4}$O$_{8}$
(Pr124) with double chains   is  the Pr$_{2}$Ba$_{4}$Cu$_{7}$O$_{15-\delta}$ (Pr247) compound 
with an alternative repetition of the CuO single chain and double chain blocks along the $c$-axis,
isostructural with superconducting Y$_{2}$Ba$_{4}$Cu$_{7}$O$_{15-\delta}$ (Y247). \cite{BO88,YA94} 
In contrast to Y123, Y247 retains its orthorhombic structure and its superconducting properties, even after oxygen along the single chains is completely depleted.\cite{GE92} This finding is considered to be due to the presence of the double CuO chain block in Y247. 
For Pr247, it is possible to examine physical properties of metallic double chains 
in terms of varying oxygen contents along single chains.

Recently, Matsukawa et al., discovered that oxygen removed polycrystalline Pr$_{2}$Ba$_{4}$Cu$_{7}$O$_{15-\delta }$   with CuO metallic double chain  shows superconductivity around $T_{c,on}=\sim 15$ K. \cite{MA04} 
Matsushita et al., revealed that Hall coefficient of oxygen defect Pr247 is varied from positive sign to negative one upon decreasing temperatures, indicating an electron doped superconductor.\cite{MA07}
The nuclear quadrapole resonance experiment indicated that the superconductivity of reduced Pr247 is realized at the CuO double chains.\cite{WA05} 
For the appearance of Pr247 superconductor, it is essential to remove oxygen from the as sintered sample of Pr247.\cite{MA04,YA05} Through several works on synthesis of higher $ T_{c}$ samples in nominal Pr247 composition, it has been made clear that the heterogeneous structure containing both Pr123 and Pr124 phases in Pr247 system plays a crucial role on attaining higher $ T_{c}$. \cite{HA07,HA08} 
High-resolution transmission electron microscopy (TEM) analyses of the heterogeneous Pr247 sample  exhibits that there exists an irregular long-period stacking structure along the $c$ axis such as \{-S-S-S-S-D-\} sequence, where S and D denote single CuO chain and double CuO chain blocks, respectively.\cite{HA07}

In this paper, we have reported the magnetic, electronic and thermal transport properties of the oxygen removed  Pr$_{2}$Ba$_{4}$Cu$_{7}$O$_{15-\delta }$. In particular, the  $\delta =0.94$ sample  exhibits higher superconductivity attaining $T_{c,on}=\sim 30$ K. 

\section{EXPERIMENT}
\begin{figure}[ht]
\includegraphics[width=8cm]{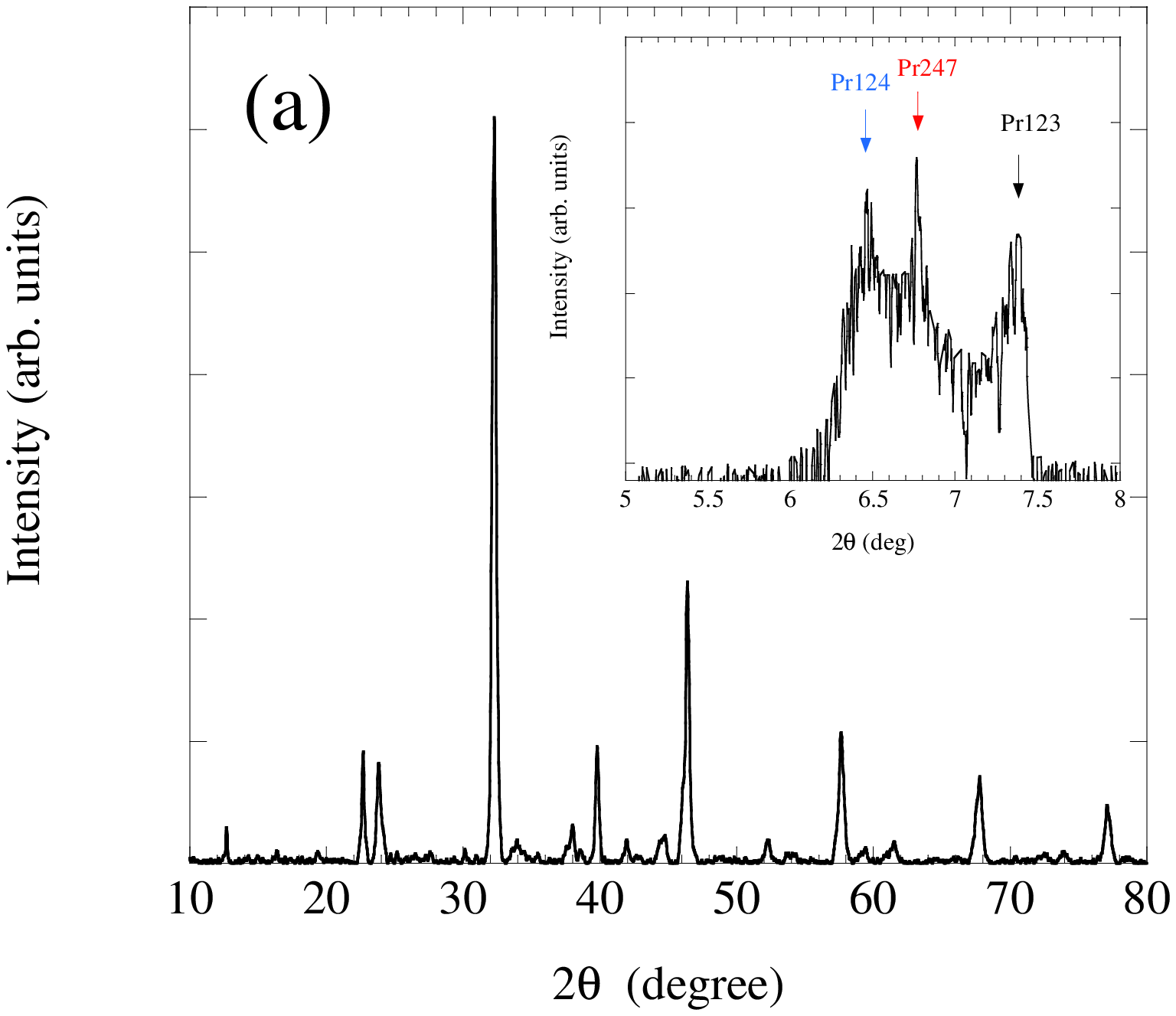}
\includegraphics[width=8cm]{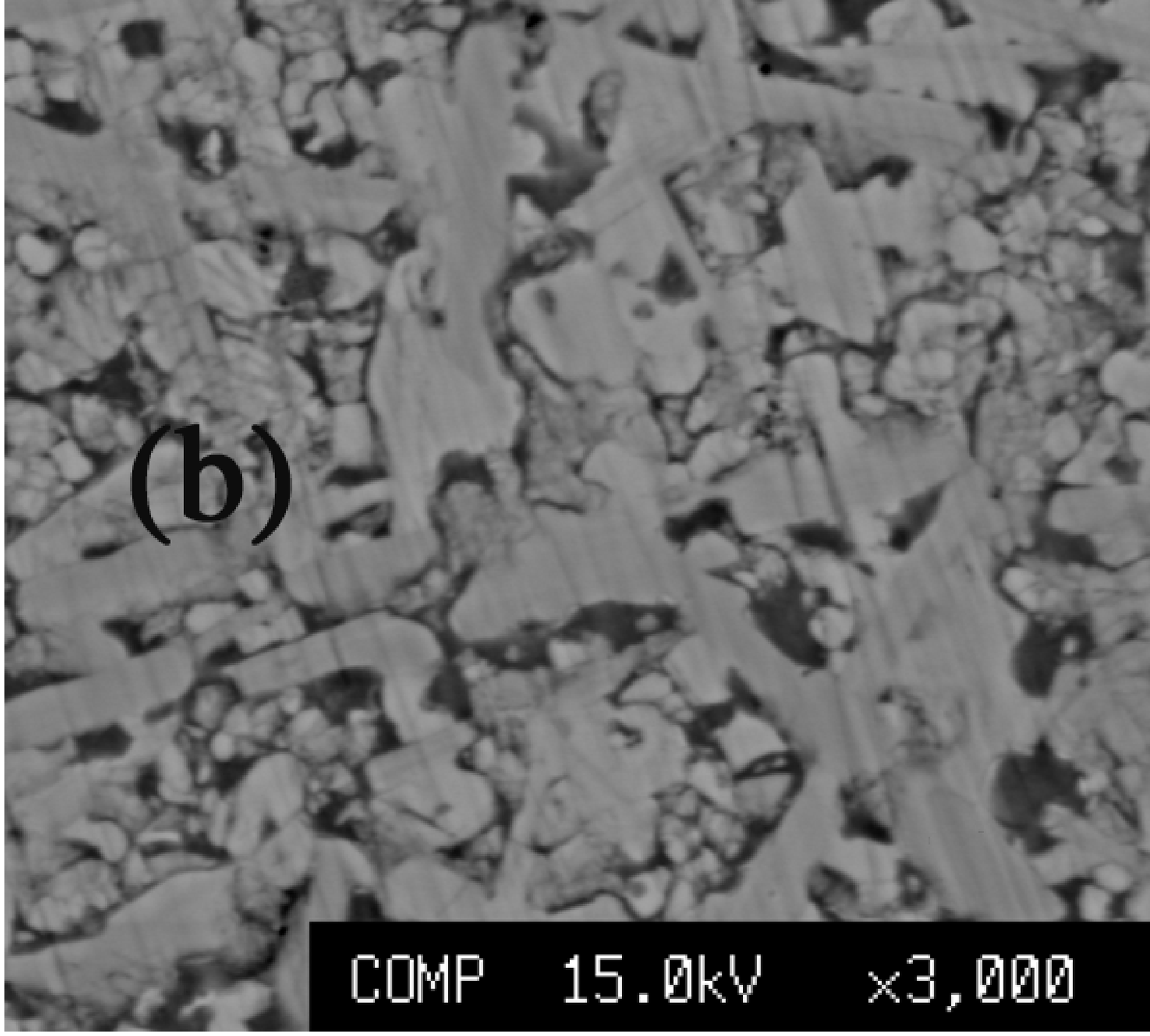}
\caption{(color online) (a)X-ray powder diffraction pattern and (b) SEM image of the superconductive sample with the nominal Pr$_{2}$Ba$_{4}$Cu$_{7}$O$_{15-\delta }$. The inset of (a) shows the low angle diffraction data enlarged near $2\theta = 5^{ \circ }-8^{ \circ }$. Here, three arrows point to the peaks corresponding to the Miller indexes (002), (004), and (001) of Pr124, Pr247, and Pr123, respectively upon increasing $2\theta $.
 }\label{xray}
\end{figure}



Polycrystalline samples of the nominal composition Pr$_{2}$Ba$_{4}$Cu$_{7}$O$_{15-\delta }$(Pr247) were synthesized using sol-gel technique. High purity powders of praseodymium oxide (Pr$_{6}$O$_{11}$), barium carbonate (BaCO$_{3}$), and  copper carbonate (CuCO$_{3}$) were mixed to the stoichiometric composition, they were then added to a suitable amount  of citric acid solution and finally their mixture was magnetically stirred in a container. The solution was condensed on a hot plate at 80 $ ^{ \circ }$C for 30 min and the resultant citrate gel  was pyrolyzed at 450 $ ^{ \circ }$C for 2 h, to remove the organic compounds. The obtained precursor was ground into a powder by a glass rod. The precursor was calcined in air at 800 $ ^{ \circ }$C for 24 h. The resultant powder was pressed into a  pellet and then sintered at 875$ \sim $876 $ ^{ \circ }$C for a long time over 180 h under ambient oxygen pressure (as-sintered sample). For the sintered process, the electric furnace equipped with three temperature controllers was used to reduce thermal gradient along the axis of furnace tube. The resultant sample was post annealed in vacuum at 500 $ ^{ \circ }$C for 48$\sim  $72 h (reduced sample). 
X-ray powder diffraction measurement was carried out with Cu$K\alpha  $ radiation at diffraction angle range of $2\theta = 5^{ \circ } - 80^{ \circ }$. 
The electron probe micro analyzer (EPMA) analysis on the superconductive sample annealed in vacuum revealed that the atomic ratio of Pr:Ba:Cu = 1.99:4.18:6.72, through an average of composition analyses at several points within its sample. The oxygen deficiency was estimated to be $\delta = 0.70$ and 0.94 from gravimetric analysis for 48 h and 72 h reduction treatments,respectively. 
The electric resistivity was measured with dc four terminal method. Seebeck coefficient $S(T)$ was determined from both measurements of a thermoelectric voltage and temperature difference along the longitudinal direction of the measured sample.  The Seebeck coefficient of copper lead was subtracted from the measured values using the published data. 
The thermal conductivity data were collected with a conventional heat flow method.  
The dc magnetization was performed using the commercial superconducting quantum interference  device (SQUID) magnetometers both at Iwate University and National Institute for Materials Science.  In particular, to remove the remanent magnetic field below 1 mOe, the fluxgate device was utilized in the zero-field cooled (ZFC) and field cooled (FC) magnetization  measurements under $H=$ 1 Oe.
Hydrostatic pressure in magnetization was applied by using a clamp-type CuBe cell up to 1.2 GPa. Fluorinert was used as a pressure transmitting medium.  The magnitude of pressure was calibrated by the pressure dependence of the critical temperature of lead.

\begin{figure}[ht]
\includegraphics[width=10cm]{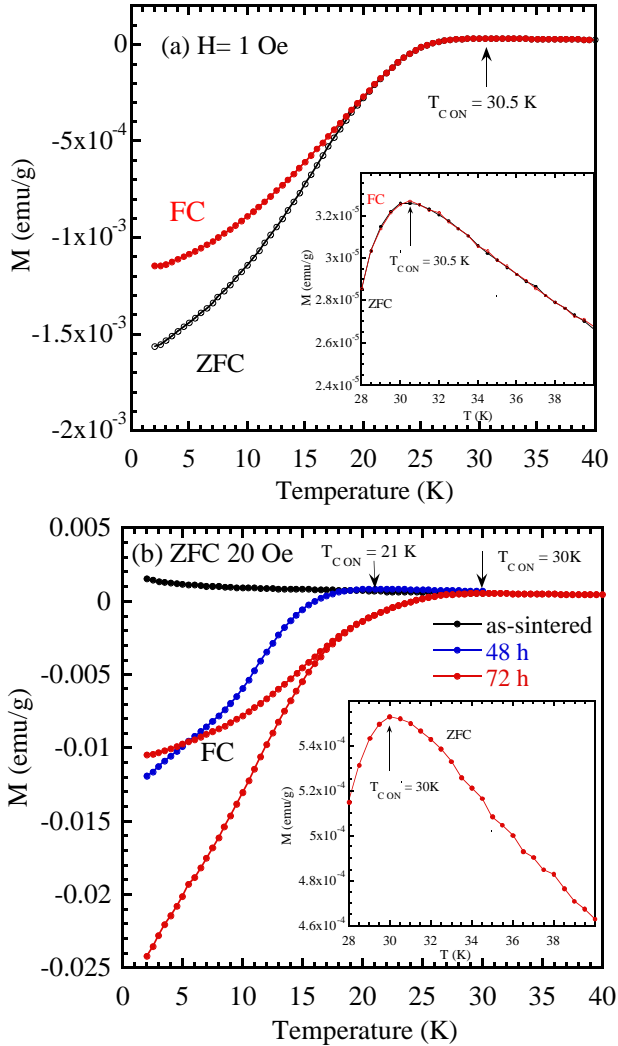}
\caption{(color online) (a)Temperature variation of zero field cooled (ZFC) and field cooled (FC) magnetization of oxygen removed Pr247 under an applied field of 1 Oe using the fluxgate device. (b) The magnetization curves of the oxygen removed Pr247 under 20 Oe for several reduced conditions. In the inset, the magnetization data are magnified to clarify the definition of  $T_{c,on}$.  
 }\label{MT}
\end{figure}%

\begin{figure}[ht]
\includegraphics[width=10cm]{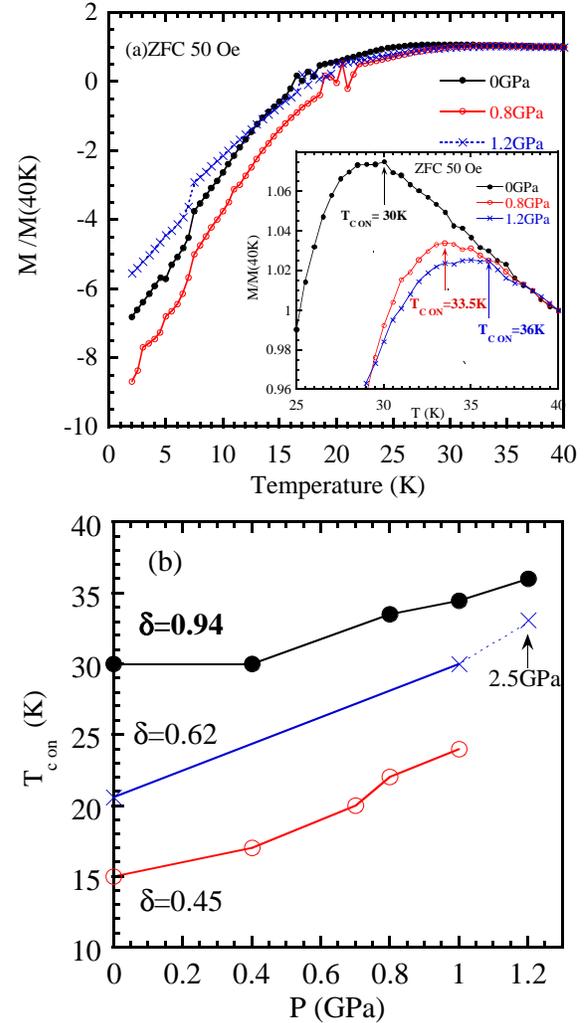}
\caption{(color online) (a) Temperature variation of normalized magnetization of oxygen removed Pr247 under hydrostatic pressures of 0, 0.8 and 1.2GPa. The enlarged plots are shown in the inset, to clarify the effect of external pressure on   $T_{c,on}$. (b) The pressure dependence of $T_{c,on}$ of the Pr247 sample with the oxygen defect $\delta = 0.94$ as a function of the applied pressure up to 1.2 GPa. For comparison, the data of pure Pr247 with $\delta = 0.45$ and 0.62 prepared by a high oxygen pressure  technique are cited.\cite{IS07,IS09}    
 }\label{MTP}
\end{figure}%
\begin{figure}[ht]
\includegraphics[width=8cm]{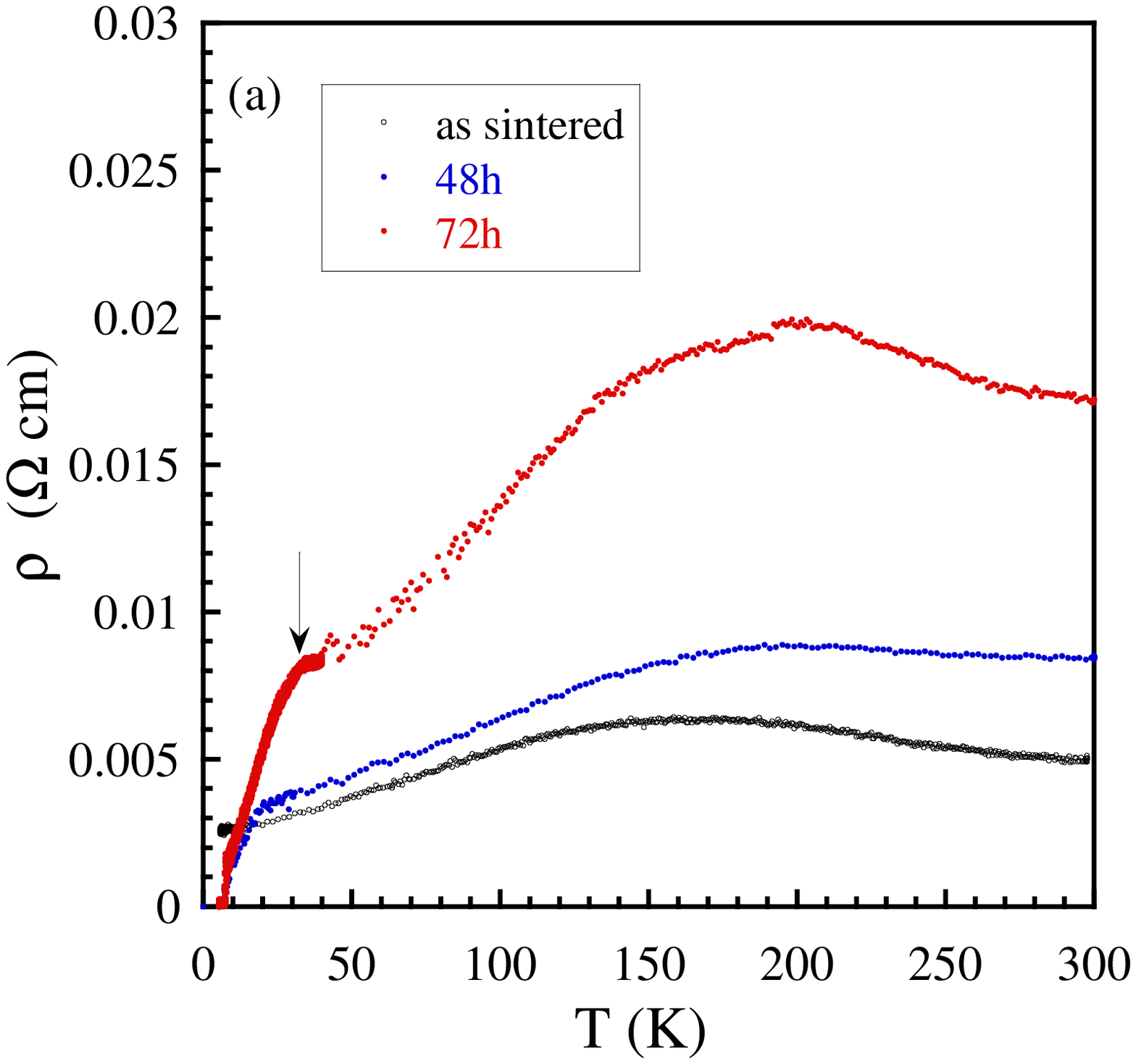}
\includegraphics[width=8cm]{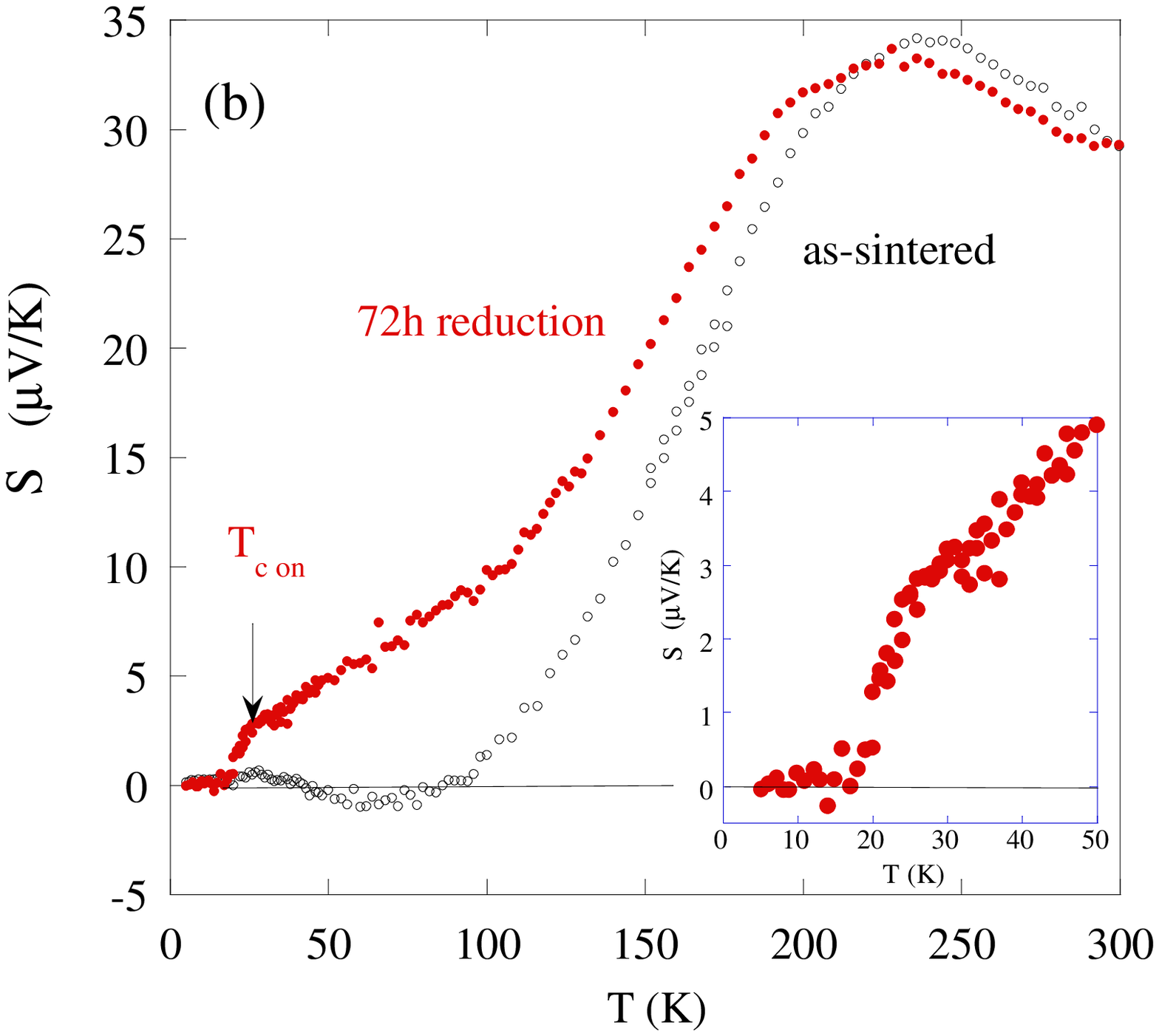}
\caption{(color online) (a)Temperature dependence of the resistivity of oxygen removed Pr247 for several annealing conditions.(b)Temperature dependence of Seebeck coefficient for the as-sintered and oxygen removed Pr247. In the inset, the magnified data of the reduced sample are presented. 
}\label{RT}
\end{figure}

\section{RESULTS AND DISCUSSION}
First of all, let us show in Fig.\ref{xray} X-ray powder diffraction pattern of the superconductive sample with the nominal Pr$_{2}$Ba$_{4}$Cu$_{7}$O$_{15-\delta }$.  The low angle diffraction data enlarged near $2\theta = 5^{ \circ }-8^{ \circ }$ are presented in the inset of Fig.\ref{xray}(a), where three arrows point to the peaks corresponding to the Miller indexes of Pr124, Pr247, and Pr123 upon increasing $2\theta $.  These findings indicate that the nominal phase of Pr247 obtained coexists with both Pr124 and Pr123 phases. Under ambient pressure condition, it seems to be difficult to synthesize high quality polycrystalline sample with nominal composition. However, to our knowledge, the heterogeneous samples have shown  good superconductive properties in comparison to pure Pr247 with lower $T_{c,on}=\sim 15$ K.\cite{HA07}  The scanning electron microscope (SEM) micrographs  of the superconducting sample with  $\delta = 0.94$ displayed in Fig.\ref{xray}(b) reveal that its typical grain size  is estimated to be  several $\mu $m. 

Figure \ref{MT} (a) shows the temperature variation of zero field cooled (ZFC) and field cooled (FC) magnetization of  the oxygen removed Pr247 sample under an applied field of 1 Oe.  In the inset of Fig. \ref{MT}, the magnified data are plotted, to clarify the definition of onset $T_{c}$.    
The as-sintered sample exhibits no superconducting anomaly as previously reported. However, the reduction treatment on its as-sintered one in vacuum results in diamagnetic character from $T_{c,on}=\sim 21$ K for 48 h up to  $T_{c,on}=\sim 30 K$ K for 72 h, which reaches highest value over the previous works. \cite{MA04,YA05,HA07} 
The magnetic susceptibility data taken under 1 Oe using the fluxgate device are presented for the 72 h reduction sample. 
From the ZFC magnetization at 2 K,  the superconducting volume fraction is estimated to be $\sim 10 \%$, indicating bulk superconductivity. 
Next, let us examine the effect of external pressure on the low-T magnetization in the oxygen removed Pr247.  First of all, application of hydrostatic pressure on the superconducting sample enhances $T_{c,on}=\sim 30$ K  at 0 GPa up to 36 K at 1.2 GPa as shown in Fig. \ref{MTP}, where the data included the magnetization of the pressure cell are normalized by the value of magnetization at 40 K.
 In the inset of Fig. \ref{MTP}(a),  the enlarged plots are displayed to emphasize the variation of $T_{c,on}$ with pressure. 
In Fig.\ref{MTP} (b), the effect of pressure on $T_{c,on}$ of pure Pr247 with $\delta = 0.45$ prepared by a high pressure oxygen technique cited from ref.\cite{IS07} shows a rapid rise of  $T_{c,on}$ from 15 K at ambient pressure up to 24 K at 1 GPa, indicating the stronger dependence of  $T_{c,on}$ on pressure in comparison to the present result. 
Recently, Ishikawa has reported that the application of pressure on the onset of $T_{c}$  in  Pr$_{2}$Ba$_{4}$Cu$_{7}$O$_{15-\delta }$ ($\delta$=0.62) increases from 20.6 K at 0 GPa up to 33.1K at 2.5 GPa. Furthermore, increasing pressure above 3 GPa, the superconductivity disappeared and under high pressure above 5 GPa the temperature variation of the resistivity showed a semiconducting like behavior.\cite{IS09}
These findings suggest that if the  pure Pr247 sample is located in the underdoped region, the present Pr247 sample is near the optimum one.   
Second, the magnitude of negative magnetization (the volume of the superconducting state) is suppressed with the applied pressure of 1.2 GPa, which is contrast to the pressure effect of $T_{c,on}$.  
Sano et al discussed the suppression of the superconductivity in the CuO double chain under pressure  on the basis of Tomonaga-Luttinger liquid theory.\cite{SA05} 
If the volume shrinkage due to the application of pressure gives rise to a decrease of the distance between the two chains of a CuO double chain, we then expect that an increase of the hopping term between the nearest neighbor oxygen sites occupying 2$p$ orbitals, resulting in a suppression of the superconducting corerelation. Their calculation does not take into account for the pressure dependence of carrier concentration, giving no explanation for the overall change of the superconducting state with pressure. The $\delta$ dependence of $T_{c,on}$ shows a monotonous increase from 15 K at $\delta=0.45$ up to 30 K at $\delta=0.94$, which is qualitatively in good agreement with a theoretical prediction by Nakano.\cite{NA07} Comparing the effect of pressure on $T_{c,on}$ with the $\delta $  dependence of it, we expect that the application of pressure gives rise to an increase in the superconducting carrier concentration. 

For several reduction conditions, the temperature variation of electrical resistivity of the present sample Pr247 is given in  Fig. \ref{RT} (a). For 72h reduced case,  $ \rho $(T) curve exhibits its dramatic drop near 30 K followed by zero resistance at low temperatures. We notice that the resistivity in the normal state for except lower temperature region rises gradually with increasing the reduction time. This tendency seems to be not anomalous  but common behavior in the present system because an increase of the resistivity in CuO$_{2}$ plane, $\rho _{pl}$, due to the reduction treatment  is predominant over a decrease of that along CuO double chain, $\rho _{ch}$ at higher temperatures. 

Finally, we check thermal transport properties of the superconducting and non-superconducting Pr247 (Fig. \ref{RT}(b)).(Thermal conductivity data are not shown here) Seebeck coefficient  $S$ is a sensitive probe to examine the electronic state of highly inhomogeneous materials since the thermal current is hard to be disturbed by the grain boundaries in contrast to the electric current.   
Upon decreasing $T$, Seebeck coefficient $S(T)$ of the superconducting sample reaches a maximum peak, then shows a rapid decrease below 200 K, and finally followed by a substantial drop below $T_{c on}$. 
It is true that the behavior of $S(T)$ of the as-sintered non-superconducting sample is, both in its temperature dependence and magnitude, similar to that of Pr124 with metallic double chain.\cite{TE96}  The magnitude of $S$ of Pr124  reaches a small value less than 1 $\mu$V/K below 100 K, which is explained on the basis of the one- to two-dimensional (1D-2D) crossover.
For conventional metal, Seebeck coefficient is expressed in terms of  $S=(\pi k)^2T/3e
\{\partial $ln$(N(E))/\partial E \}$, where $N(E)$ is the density of electronic state.\cite{ZI76} In our rough estimation, if we assume that the magnitude of S is proportinal to the energy derivative of  $N(E)$,  the density of state for 2D metal is independent of the energy, giving $S(T)\sim 0$.
However, the $S$ value of the superconducting Pr247 shows a finite value of at most a few $\mu$V/K  down to $T_{c on}$. This finding is probably ascribed to a variation of the electronic state of the CuO double chain associated with carrier doping into the double-chain block due to the reduction treatment. 
Upon decreasing $T$, the Hall coefficient $R_{H}$ of the 24 h reduced sample  changes its sign across 100 K and then exhibits a rapid increases with negative sign.  We expect that the disagreement in sign between  $R_{H}$ and $S$ in the oxygen removed sample is ascribed to the dramatic variation of the density of states with reduction treatment since $\partial (N(E))/\partial E $ near the Fermi energy is related to the sign of Seebeck coefficient in addition to the sign of charge carrier. 
 
 
A previous work\cite{LO90} on role of oxygen in Pr123 compound revealed that  the oxygen deficient Pr123 phase remains highly insulating state, causing neither metallic or superconducting sign. 
Double chain conduction in single crystalline Pr124 exhibits a strongly metallic character but no superconducting transition down to 2 K.\cite{HO00} 
We concentrate on single-chain rich microdomains present in the nominal Pr247 as observed by TEM measurements\cite{HA07}, which is probably close to the appearance of higher $T_{c,on}$= 30 K accompanied by the larger oxygen defect $\delta = 0.94$.  
It is, thus, expected that a reduction treatment removes oxygen from CuO single chain rich sequence such as \{S-D-S-S-S-D\} and results in  higher carrier doping  into its surrounding CuO double chain in comparison to that in the regular sequence \{S-D-S-D-S-D\}.  

In summary, we have demonstrated the pressure effect of $ T_{c,on}$, and the transport properties of the nominal composition Pr$_{2}$Ba$_{4}$Cu$_{7}$O$_{15-\delta }$. Magnetization and resistivity measurements of the $\delta$=0.94 sample reduced in vacuum exhibit $T_{c,on}=\sim 30$ K. 	
The pressure dependence of $T_{c,on}$ of the $\delta = 0.94$  sample strongly suggests the nearly optimum doping of the superconducting carries. Comparing the effect of pressure on $T_{c,on}$ with the $\delta $  dependence of it, we notice that the application of pressure gives rise to an increase in the carrier concentration. 
We believe that the heterogeneous structure of the nominal Pr247 is closely related to the appearance of higher $T_{C}$.  Seebeck coefficient of the $\delta$=0.94 sample shows a metallic conduction below 100 K and is in its temperature dependence quite different from the data of the as-sintered one, leading to a dramatic change of the density of state along the CuO metallic double chain. 

\begin{acknowledgments}
The authors are grateful for Dr. K. Nonaka for his assistance in EPMA experiments.
This work was partially supported by a Grant-in-Aid for Scientific Research from Japan Society of the Promotion of Science. 
\end{acknowledgments}

\end{document}